\documentclass{article} 
\usepackage{epsfig} 
\usepackage{cite} 
\def\H{\hbox{H}}

\def\f{\hbox{f}}

\def\ln{\hbox{ln}} 
\def\hpl{HPL} 
\def\ie{{\it i.e.}\phantom{.}}

\begin{document} 
\unitlength1cm 
\begin{titlepage} 
\vspace*{-1cm} 
\begin{flushright} 
TTP02-01\\ 
%%hep-ph/0202123\\ 
February 2002 
\end{flushright} 
\vskip 3.5cm 
\renewcommand{\topfraction}{0.9}
\renewcommand{\textfraction}{0.0}

\begin{center} 
\boldmath 
{\Large\bf The analytic value of the sunrise self-mass with two equal 
masses and the external invariant equal to the third squared mass. 
}\unboldmath 
\vskip 1.cm 
{\large M. Argeri}$^a$, {\large P. Mastrolia}$^{a,b}$, and 
{\large E. Remiddi}$^{a,c}$ 
\vskip .7cm 
% {\it $^a$ Theory Division, CERN, CH-1211 Geneva 23, Switzerland} 
\vskip .4cm 
{\it $^a$ Dipartimento di Fisica, 
    Universit\`{a} di Bologna, I-40126 Bologna, Italy \\
       $^b$ Institut f\"ur Theoretische Teilchenphysik,
            Universit\"at Karlsruhe, D-76128 Karlsruhe, Germany \\ 
       $^c$ INFN, Sezione di Bologna, I-40126 Bologna, Italy \\ 
} 
\end{center} 
\vskip 2.6cm 

\begin{abstract} 
We consider the two-loop self-mass sunrise amplitude with two equal 
masses $M$ and the external invariant equal to the square of the 
third mass $m$ in the usual $d$-continuous dimensional regularization. 
We write a second order differential equation for the amplitude 
in $x=m/M$ and show as solve it in close analytic form. As a result, 
all the coefficients of the Laurent expansion in $(d-4)$ 
of the amplitude are expressed in terms of harmonic polylogarithms of 
argument $x$ and increasing weight. As a by product, we give 
the explicit analytic expressions of the value of the amplitude at 
$x=1$, corresponding to the on-mass-shell sunrise amplitude in the 
equal mass case, up to the $(d-4)^5$ term included. 
\end{abstract} 
\vfill 
\end{titlepage} 
\newpage 
\renewcommand{\theequation}{\mbox{\arabic{section}.\arabic{equation}}} 
\section{Introduction} 
\label{sec:int} 
\begin{figure}[h] 
\begin{center} 
\includegraphics*[2cm,1cm][14cm,6cm]{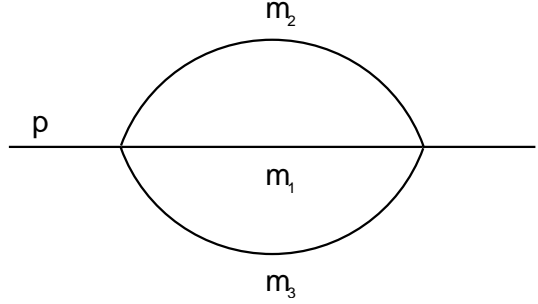} 
\caption{The sunrise graph for arbitrary masses.} 
\end{center}
\label{fig:sunarb} 
\end{figure}

\setcounter{equation}{0} 
It is known that the sunrise graph of Figure~\ref{fig:sunarb} in the 
arbitrary mass case has four master integrals; 
%%%%%%%%%%%%%%%%%%%%%%%%%%%%%%%%%%%%%%%%%%%%%%%%%%%%%%%%%%%%%%%%%% 
%%%%%%%%%%%%%%%%%%%%%%%%%%%%%%%%%%%%%%%%%%%%%%%%%%%%%%%%%%%%%%%%%% 
they can be taken to be the fully scalar amplitude 
\begin{equation} 
 F_0(d,m_1^2,m_2^2,m_3^2,p^2) = \int \frac{d^dk_1}{(2\pi)^{d-2}} 
                                     \frac{d^dk_2}{(2\pi)^{d-2}} 
    \frac{1}{ [(p-k_1)^2+m_1^2][(k_1-k_2)^2+m_2^2](k_2^2+m_3^2) }, 
\end{equation} 
where $d$ is the continuous dimension, $m_i, i=1,3$ the masses, $p^2$ 
the square of the external momentum (all momenta being Euclidean), 
and the three amplitudes $F_i$ defined as 
\begin{equation} 
    F_i(d,m_1^2,m_2^2,m_3^2,p^2) = 
    - \frac{\partial}{\partial m_i^2} F_0(d,m_1^2,m_2^2,m_3^2,p^2) \ , 
\end{equation} 
for $i=1,3$. The higher mass derivatives of the amplitudes can all be 
expressed, as a consequence of the integration by parts identities, 
in terms of the four master integrals. 
The four master integrals satisfy further a system of linear non homogeneous 
differential equations in $p^2$, derived in \cite{sreq}. 
\par 
In the particular kinematical configuration $ - p^2 = m_1^2 = m^2 $ 
and $ m_2^2 = m_3^2 = M^2 $, Figure~2, one has 
%%%%%%%%%%%%%%%%%%%%%%%%%%%%%%%%%%%%%%%%%%%%%%%%%%%%%%%%%%%%%%%%%% 
\begin{figure}[h]
\begin{center} 
\includegraphics*[2cm,1cm][14cm,6cm]{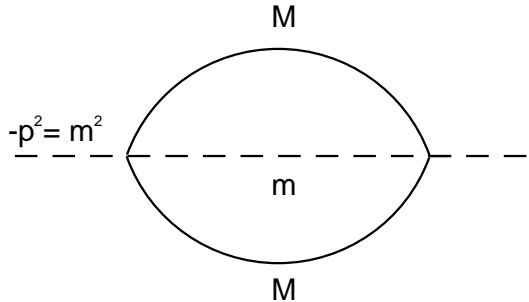} 
\caption{The sunrise graph with two equal masses and the external invariant 
equal to the third squared mass.} 
\end{center}
\label{fig:sunpart}
\end{figure}
%%%%%%%%%%%%%%%%%%%%%%%%%%%%%%%%%%%%%%%%%%%%%%%%%%%%%%%%%%%%%%%%%% 
\begin{equation} 
    F_0(d,m^2,M^2,M^2,p^2=-m^2) = \int \frac{d^dk_1}{(2\pi)^{d-2}} 
                                       \frac{d^dk_2}{(2\pi)^{d-2}} 
    \frac{1}{ (k_1^2-2pk_1)[(k_1-k_2)^2+M^2](k_2^2+M^2) } \ ; 
\label{eq:F0part} 
\end{equation} 
introducing the dimensionless variables $ x = m/M $ one can define the 
dimensionless function $\Phi(d,x)$ as 
\begin{equation} 
 F_0(d,m^2,M^2,M^2,-m^2) = M^{2d-6} C^2(d) \Phi(d,x) \ , 
\label{eq:defPhi1} 
\end{equation} 
{\it i.e.} 
\begin{equation} 
 \Phi(d,x) = C^{-2}(d) M^{6-2d} F_0(d,M^2x^2,M^2,M^2,-M^2x^2) \ , 
\label{eq:defPhi2} 
\end{equation} 
where $C(d)= (4\pi)^\frac{4-d}{2}\Gamma(3-d/2) $ is an overall normalization 
factor, with the limiting value $C(4)=1$ at $d=4$. \par 
It will be shown in this paper that the integration by part identities for the 
master integrals imply a second order differential equation in $x$ 
for $\Phi(d,x)$; when functions and equations are systematically expanded 
around $d=4$ as a Laurent series in $(d-4)$, one obtains a system of chained 
differential equations which can be easily solved recursively in closed 
analytical form, virtually up to any order in $(d-4)$; one finds in this 
way that the coefficients of the $(d-4)$ expansion of $\Phi(d,x)$ 
are a combination of harmonic polylogarithms~\cite{hpl} of increasing 
weight. \par 
The differential equations themselves with the knowledge of the behaviour 
at $m=0$ can be used to fix completely the integration constants; as a 
byproduct we give the value of the integral at $m=M$ (on mass shell equal 
mass case) up to the order $(d-4)^{5}$ included. 

\section{The differential equations.} 
\label{sec:diffeq} 
\setcounter{equation}{0} 
For deriving the differential equation for $\Phi(d,x)$ we 
introduce the auxiliary functions 
\begin{eqnarray} 
 \Psi_0(d,\eta,p^2) &=& m^{6-2d}\ F_0(d,m^2,m^2\eta,m^2\eta,p^2) \ , 
\label{eq:defPsi3} \\ 
 \Psi_1(d,\eta,p^2) &=& m^{8-2d}\ \bigl[ F_2(d,m^2,m^2\eta,m^2\eta,p^2) 
                               + F_3(d,m^2,m^2\eta,m^2\eta,p^2) \bigr] \ . 
\end{eqnarray} 
One finds at once 
\begin{equation} 
 \frac{\partial}{\partial \eta}\Psi_0(d,\eta,p^2) = - \Psi_1(d,\eta,p^2) \ , 
\end{equation} 
from which in turn 
\begin{eqnarray} 
 \frac{\partial}{\partial \eta}\Psi_1(d,\eta,p^2) &=& - m^{(10-2d)} 
                                                              \nonumber\\ 
         && \left(   \frac{\partial}{\partial m_2^2} 
                   + \frac{\partial}{\partial m_3^2} \right) 
      \bigl[ F_2(d,m^2,m_2^2,m_3^2,p^2) + F_3(d,m^2,m_2^2,m_3^2,p^2) 
      \bigr]_{m_2^2=m_3^2=m^2\eta} \ . 
\end{eqnarray} 
As already remarked, the mass derivatives of the master integrals 
$F_i, i=1,3$ can be expressed, by using the integration by parts identities, 
in terms of the master integrals themselves (see for instance the Appendix 
of ~\cite{sreq}), so that the above quantity is equal to a combination 
of $F_0$, corresponding to $ \Psi_0(d,\eta,p^2) $, of the sum 
$F_2+F_3$ (as required by the $m_2=m_3$ symmetry), corresponding to 
$ \Psi_1(d,\eta,p^2) $, and of the fourth master integral $F_1$, independent 
of $ \Psi_0 $ and $ \Psi_1 $, plus an inhomogeneous part given 
by a few products of tadpoles $T(d,m^2), T(d,M^2)$, with 
\begin{equation} 
  T(d,m^2) = \int \frac{d^dk}{(2\pi)^{d-2}} \frac{1}{k^2+m^2} = 
             C(d) \; \frac{m^{d-2}}{(d-2)(d-4)} \ . 
\end{equation} 
The coefficients of the various terms are ratios of polynomials 
in $d$, $p^2$ and the masses; in the limit $p^2\to -m^2$ the 
denominators of the coefficients develop a singular $1/(p^2+m^2)^2$ behaviour, 
which is however fully compensated by corresponding zeroes in the numerators, 
so that the limit $p^2 = -m^2$ is finite. In that limit, further, the 
coefficient of the master amplitude $F_1$ vanishes, so that 
$ \frac{\partial}{\partial \eta}\Psi_1(d,\eta,p^2=-m^2) $, which is 
essentially the second $\eta$-derivative of $ \Psi_0(d,\eta,p^2=-m^2) $, 
is expressed in terms of $\Psi_1(d,\eta,p^2=-m^2) $, its first 
$\eta$-derivative (up to a sign), and of $ \Psi_0(d,\eta,p^2=-m^2) $ 
itself, plus an inhomogeneous term given by the tadpoles. 
For taking the $p^2 = -m^2$ limit, let us introduce the function 
$ \Psi(d,\eta) $ through 
\begin{equation} 
 F_0(d,m^2,M^2,M^2,-m^2) = m^{2d-6}\ C^2(d) \ \Psi(d,\eta) \ , 
\label{eq:defPsi1} 
\end{equation} 
{\it i.e.} 
\begin{equation} 
 \Psi(d,\eta) = C^{-2}(d)\ m^{6-2d}\ F_0(d,m^2,m^2\eta,m^2\eta,-m^2) \ . 
\label{eq:defPsi2} 
\end{equation} 
In terms of $ \Psi(d,\eta) $ the second order differential equation for 
$ \Psi_0(d,\eta,p^2=-m^2) $ then reads 
\begin{eqnarray} 
   \eta(1-\eta)\frac{\partial^2}{\partial \eta^2} \Psi(d,\eta) 
  - \left[ \frac{1}{2} - \frac{3}{2}\eta + (1-2\eta)(d-4) \right] 
    \frac{\partial}{\partial \eta} \Psi(d,\eta) && \nonumber\\ 
  - \left[ 1 + \frac{7}{4}(d-4) + \frac{3}{4}(d-4)^2 \right] \Psi(d,\eta) 
  &=& \frac{\eta^{(d-4)/2}}{4(d-4)^2} 
    - \frac{\eta^{1+(d-4)}}{8(d-4)^2} \ . 
\label{eq:eqeta} 
\end{eqnarray} 
The equation, exact in the continuous dimension $d$, has been written in 
terms of powers of $(d-4)$ for convenience of later use. 
\par 
As $\eta = 1/x^2 $, by comparing the definitions 
Eq.s~(\ref{eq:defPhi1}, \ref{eq:defPsi1}) one has 
\begin{equation} 
  \Psi\left(d,\eta = \frac{1}{x^2}\right) = x^{6-2d} \Phi(d,x) \ , 
\label{eq:etatox} 
\end{equation} 
and the above equation for $\Psi(d,\eta)$ can be converted in the following 
equation for $\Phi(d,x)$ 

\begin{eqnarray} 
  \frac{1}{4}x^2(x^2-1) \frac{\partial^2}{\partial x^2} \Phi(d,x) 
 -\frac{1}{2}x \left[ 1 + (d-4)x^2 \right] 
  \frac{\partial}{\partial x} \Phi(d,x) && \nonumber\\ 
  + \left[ \frac{1}{2}(1 - x^2 ) 
           + \left( \frac{3}{4} - \frac{1}{2}x^2 \right) (d-4) 
           + \frac{1}{4} (d-4)^2 \right] \Phi(d,x) 
  &=& \frac{x^{2+(d-4)}}{4(d-4)^2} - \frac{1}{8(d-4)^2} \ . 
\label{eq:eqx} 
\end{eqnarray} 

\section{The expansion in $(d-4)$ around $d=4$ . } 
\label{sec:dexp} 
\setcounter{equation}{0} 
It is known~\cite{sreq} that in the $d\to 4$ limit $\Phi(d,x)$ develops a 
double pole in $(d-4)$, so that its Laurent expansion in $(d-4)$ reads 
\begin{equation} 
  \Phi(d,x) = \sum_{n=-2}^\infty (d-4)^n \Phi^{(n)}(x) \ . 
\label{eq:Phidexp} 
\end{equation} 
By inserting the above expansion in Eq.~(\ref{eq:eqx}) one obtains a system 
of chained inhomogeneous equations in the functions $ \Phi^{(n)}(x)$. 
For $n=-2,-1$ they read 
\begin{eqnarray} 
  \left[                       \frac{\partial^2}{\partial x^2} 
  + \left( \frac{2}{x} + \frac{1}{1-x} - \frac{1}{1+x} \right) 
                                   \frac{\partial}{\partial x} 
  - \frac{2}{x^2} \right] \Phi^{(-2)}(x) = && 
  \frac{1}{2x^2} - \frac{1}{4(1-x)} - \frac{1}{4(1+x)} \nonumber\\ 
  \left[                       \frac{\partial^2}{\partial x^2} 
  + \left( \frac{2}{x} + \frac{1}{1-x} - \frac{1}{1+x} \right) 
                                   \frac{\partial}{\partial x} 
  - \frac{2}{x^2} \right] \Phi^{(-1)}(x) = 
  &-& \left(  \frac{1}{1-x} - \frac{1}{1+x} \right) 
            \frac{\partial}{\partial x} \Phi^{(-2)}(x) \nonumber\\ 
  + \left( \frac{3}{x^2} + \frac{1}{2(1-x)} +  \frac{1}{2(1+x)} 
    \right)  \Phi^{(-2)}(x) 
  &-& \frac{1}{2}\left( \frac{1}{1-x} +  \frac{1}{1+x} \right) \ln{x} \ , 
\label{eq:exp-2-1} 
\end{eqnarray} 
while for $ n \ge 0 $ one has 
\begin{eqnarray} 
  \left[                       \frac{\partial^2}{\partial x^2} 
  + \left( \frac{2}{x} + \frac{1}{1-x} - \frac{1}{1+x} \right) 
                                   \frac{\partial}{\partial x} 
  - \frac{2}{x^2} \right] \Phi^{(n)}(x) = 
  &-& \left(  \frac{1}{1-x} - \frac{1}{1+x} \right) 
            \frac{\partial}{\partial x} \Phi^{(n-1)}(x) \nonumber\\ 
  + \left( \frac{3}{x^2} + \frac{1}{2(1-x)} +  \frac{1}{2(1+x)} 
    \right)  \Phi^{(n-1)}(x) 
  &+& \left( \frac{1}{x^2} + \frac{1}{2(1-x)} +  \frac{1}{2(1+x)} 
    \right)  \Phi^{(n-2)}(x)                            \nonumber\\ 
  &-& \frac{1}{2}\left( \frac{1}{1-x} +  \frac{1}{1+x} \right) 
                                     \frac{\ln^{n+2}{x}}{(n+2)!} \ . 
\label{eq:expn} 
\end{eqnarray} 
A few comments are in order. Quite in general, all the equations have 
the form 
\begin{equation} 
  \left[                       \frac{\partial^2}{\partial x^2} 
  + \left( \frac{2}{x} + \frac{1}{1-x} - \frac{1}{1+x} \right) 
                                   \frac{\partial}{\partial x} 
  - \frac{2}{x^2} \right] \Phi^{(n)}(x) = R^{(n)}(x) \ ; 
\label{eq:generic} 
\end{equation} 
the homogeneous part of the equations is always 
\begin{equation} 
  \left[                       \frac{\partial^2}{\partial x^2} 
  + \left( \frac{2}{x} + \frac{1}{1-x} - \frac{1}{1+x} \right) 
                                   \frac{\partial}{\partial x} 
  - \frac{2}{x^2} \right] \phi(x) = 0 \ , 
\label{eq:homo} 
\end{equation} 
while the inhomogeneous terms $ R^{(n)}(x) $ vary from equation to equation
and involve, besides the coefficients of the $(d-4)$-expansion of the 
inhomogeneous term of Eq.(\ref{eq:eqx}) which are explicitly known since 
the beginning, also the coefficients $ \Phi^{(n-1)}(x) $ and 
$ \Phi^{(n-2)}(x) $ of the $(d-4)$-expansion of the unknown function 
$ \Psi(d,x) $. 
However, in a systematic bottom-up approach to the solution starting from 
the smallest value of $n$, \ie $\ n=-2$, the inhomogeneous part can also be 
considered as known. \par 
Finally, all the coefficients appearing in the equation, both in the 
homogeneous and the inhomogeneous part, are simple combinations of the 
rational factors $1/x$, $1/(1-x)$ and $1/(1+x)$; powers of $\ln{x}$ are 
also present, from the expansion of the original inhomogeneous term. 

\section{The behaviour of the equation for $ x \to 0^+$.} 
\label{sec:xito0} 
\setcounter{equation}{0} 
As $x$ is proportional to a mass and therefore is an essentially real and 
positive variable, we study Eq.(\ref{eq:eqx}) in the $ x \to 0^+$ limit; 
to that aim, we try the most general expansion 
\begin{equation} 
   \Phi(d,x) = \sum_i x^{\alpha_i} 
                 \left( \sum_{n=0}^\infty a_n^{(i)}(d) x^n \right) \ . 
\end{equation} 
By substituting the above expansion in Eq.(\ref{eq:eqx}), we find for the 
leading exponents $ \alpha_i $ the four allowed values 
$ \alpha_1 = (2-d) $, $ \ \alpha_2 = (d-3) $, corresponding to the two 
independent solutions of the associated homogeneous equation and 
$ \alpha_3 = (d-2) $, $ \ \alpha_4 = 0$, as required by the inhomogeneous 
terms appearing in the {\it r.h.s.} of Eq.(\ref{eq:eqx}). \par 
By direct inspection of Eq.(\ref{eq:F0part}) and 
Eq.(\ref{eq:defPhi1}), one sees that $\Phi(d,x)$ is finite for 
$ d > 2  $ in the $ x\to0^+$ limit, so that the first two behaviours, 
with exponents $ (2-d) $ and $ (d-3) $, are ruled out ({\it i.e.} the 
coefficients $ a_n^{(i)}(d) $ are all vanishing for $i=1,2$ and any 
value of $n$ ). The coefficients of the other terms are then completely 
fixed by the inhomogeneous part of the equation, and an elementary 
calculation gives the following explicit behaviour in the $x\to0^+$ limit 
\begin{eqnarray} 
  \Phi(d,x) &=& - \frac{1}{2(d-2)(d-4)^2} \left[ \frac{1}{d-3} 
                + \frac{2 x^2}{d(d-5)} \ + x^2x^{d-4} \right] \nonumber\\ 
              &+& {\mathcal O}(x^3) \ . 
\label{eq:Phiexpl} 
\end{eqnarray} 
When expanding in $(d-4)$ according to Eq.(\ref{eq:Phidexp}), the above 
equation reads 
\begin{eqnarray} 
 \Phi^{(-2)}(x) &=& - \frac{1}{2} - \frac{1}{4}x^2 
                                      + {\mathcal O}(x^3) \ , \nonumber\\ 
 \Phi^{(-1)}(x) &=& + \frac{3}{4} + \frac{5}{16}x^2 - \frac{x^2}{2} \ln{x} 
                                      + {\mathcal O}(x^3) \ , \nonumber\\ 
 \Phi^{(0)}(x)  &=& - \frac{7}{8} + \frac{3}{64}x^2 + \frac{x^2}{4} 
                ( \ln{x} - \ln^2{x} ) + {\mathcal O}(x^3) \ , \nonumber\\ 
\label{eq:Phiexpld} 
\end{eqnarray} 
and so on up to any desired order in $(d-4)$. 
\par 
It is to be observed that by using the differential equation 
Eq.(\ref{eq:eqx}) one can 
transform a qualitative information, the finiteness of $ \Phi(d,x)$ at 
$x=0$, into a quantitative information, such as the actual value of the 
$x\to 0^+$ behaviour of $ \Phi(d,x) $ given by Eq.(\ref{eq:Phiexpl}),
without carrying out any explicit loop integration in Eq.(\ref{eq:F0part}). 

\section{The solution. } 
\label{sec:sol} 
\setcounter{equation}{0} 
The algorithm for solving Eq.(\ref{eq:generic}) goes back to Euler. Let 
$ \phi_1(x) $, $ \phi_2(x) $ be two independent solutions of the associated 
homogeneous equation (\ref{eq:homo}), such that their Wronskian $ W(x) $, 
defined as 
\begin{equation} 
   W(x) = \phi_1(x) \phi'_2(x) - \phi'_1(x)  \phi_2(x) \ , 
\label{eq:defW} 
\end{equation} 
does not vanish; the solution of the second order Eq.(\ref{eq:generic}) 
can then be written as 
\begin{equation} 
   \Phi^{(n)}(x) = \phi_1(x) \left(  \Phi^{(n)}_1 
         - \int^x \frac{dx' \phi_2(x')}{W(x')} R^{(n)}(x') \right) 
                 + \phi_2(x) \left(  \Phi^{(n)}_2 
         + \int^x \frac{dx' \phi_1(x')}{W(x')} R^{(n)}(x') \right) \ , 
\label{eq:thesol} 
\end{equation} 
where $ \Phi^{(n)}_1, \Phi^{(n)}_2 $ are two integration constants, to be 
fixed by the boundary conditions Eq.(\ref{eq:Phiexpl}) expanded in $(d-4)$, 
\ie by  Eq.s(\ref{eq:Phiexpld}). 
\par 
For the actual use of Euler's algorithm, the solutions of the homogeneous 
equation (\ref{eq:homo}) are needed. It turns out that the homogeneous 
equation associated to Eq.({\ref{eq:eqeta}) at $d=4$ 
\begin{equation} 
   \left[ \eta(1-\eta)\frac{\partial^2}{\partial \eta^2} 
  - \left( \frac{1}{2} - \frac{3}{2}\eta \right) 
    \frac{\partial}{\partial \eta} - 1 \right] \psi(\eta) = 0 \ , 
\label{eq:homoeta} 
\end{equation} 
is a hypergeometric equation, whose solution regular at $\eta = 0 $ is 
the very simple hypergeometric function 
\begin{equation} 
  \,_2F_1\left(-\frac{1}{2},-2,-\frac{1}{2};\eta\right) = (1-\eta)^2 \ . 
\label{eq:2F1} 
\end{equation} 
With that input and the transformation (\ref{eq:etatox}) it 
is easy to obtain the following solutions of the homogeneous 
equation (\ref{eq:homo}): 
\begin{eqnarray} 
  \phi_1(x) &=& \frac{(1-x^2)^2}{x^2} \ , \nonumber\\ 
  \phi_2(x) &=& \frac{(1-x^2)^2}{x^2}\ \ln\left(\frac{1+x}{1-x}\right) 
                - 2\frac{1+x^2}{x} \ , 
\label{eq:homosol} 
\end{eqnarray} 
having the Wronskian 
\begin{equation} 
  W(x) = 16 \frac{1-x^2}{x^2} \ , 
\label{eq:Wrsol} 
\end{equation} 
and the $x\to0$ behaviour 
\begin{eqnarray} 
  \phi_1(x) &\to& \frac{1}{x^2} \ , \nonumber\\ 
  \phi_2(x) &\to& - 6x \ . 
\label{eq:xto0be} 
\end{eqnarray} 
With the above explicit homogeneous solutions one can solve recursively 
the system of the non homogeneous equations (\ref{eq:expn}) for 
$\Phi^{(n)}(x)$, starting from $n=-2$, by carrying out 
repeatedly the integration in $x'$ of (\ref{eq:thesol}) 
in close analytic form; one obtains in that way 
a combination of Harmonic PolyLogarithms (HPLs) 
of increasing weight~ \cite{hpl} times simple coefficients involving the 
rational factors $1/x$, $\ 1/(1-x)$ and $\ 1/(1+x)$. 
\par 
To illustrate this statement, which is the central point of this paper, 
let us recall for convenience of the reader the definition of the \hpl's. 
The \hpl's form a family of functions depending on an argument, say $x$, 
and on a set of indices, say $a_i, i=1,..,w$ or $\vec a$ in more compact 
notation, where each of the $a_i$ can take one of the three values $1, 0, -1$ 
and whose number $w$ is called the weight of the \hpl. At weight $w=1$ 
there are 3 \hpl's, defined as 
\begin{eqnarray} 
  \H(1;x) &=& - \ln(1-x) \ , \nonumber\\ 
  \H(0;x) &=& \ln{x} \ , \nonumber\\ 
  \H(-1;x) &=& \ln(1+x) \ , 
\label{eq:defw1} 
\end{eqnarray}
whose derivatives can be written as 
\begin{equation} 
  \frac{d}{dx}\H(a;x) = \f(a,x) \ , 
\label{eq:defdHw1} 
\end{equation} 
where the index $a$ can take one of the three values $(1,0,-1)$ and the 
rational factors $\f(a;x)$ are given by 
\begin{eqnarray} 
  \f(1;x)  &=& \frac{1}{1-x} \ , \nonumber\\ 
  \f(0;x)  &=& \frac{1}{x} \ , \nonumber\\ 
  \f(-1;x) &=& \frac{1}{1+x} \ . 
\label{eq:deff} 
\end{eqnarray} 
\par 
At weight $w>1$, if all the $w$ indices are equal to $0$ let us indicate 
them by $\vec 0_w$ and define correspondingly 
\begin{equation} 
   \H(\vec 0_w;x) = \frac{1}{w!} \ln^w{x} \ ; 
\label{eq:H0w} 
\end{equation} 
in all the other cases (\ie when the indices are not all equal to zero), 
let us indicate any set of $w$ indices by 
$(a,\vec b)$, where $a$ can take one of the three values $(1,0,-1)$ and 
$\vec b$ stands for the set of other $w-1$ indices, and define correspondingly 
\begin{equation} 
   \H(a,\vec b;x) = \int_0^x dx' \f(a;x') \H(\vec b;x') \ . 
\label{eq:Hw} 
\end{equation} 
Note that in full generality one has 
\begin{equation} 
  \frac{d}{dx}\H(a,\vec b;x) = \f(a,x) \H(\vec b;x) \ , 
\label{eq:dxHw} 
\end{equation} 
which can also be written as the equivalent indefinite integration formula 
\begin{equation} 
   \int^x dx' \f(a;x') \H(\vec b;x') = A + \H(a,\vec b;x) \ , 
\label{eq:intdxHw} 
\end{equation} 
where $A$ is an integration constant. \par 
Further, the product of two \hpl's  of a same argument $x$ 
and weights $p, q$ can be expressed as a combination of \hpl's  of 
argument $x$ and weight $r=p+q$, according to the product identity 
\begin{eqnarray} 
 \H(\vec{p};x)\H(\vec{q};x) & = & 
  \sum_{\vec{r} = \vec{p}\uplus \vec{q}} \H(\vec{r};x) \; , 
\label{eq:halgebra} \end{eqnarray} 
where $\vec p, \vec q$ stand for the $p$ and $q$ components of the indices 
of the two \hpl's, while $\vec{p}\uplus \vec{q}$ represents all possible 
mergers of $\vec{p}$ and $\vec{q}$ into the vector $\vec{r}$ with $r$ 
components, in which the relative orders of the elements of $\vec{p}$ 
and $\vec{q}$ are preserved. The simplest cases of the above identities 
are 
\begin{eqnarray} 
 \H(a;x)\H(b;x) &=& \H(a,b;x) + \H(b,a;x) \ , \nonumber\\ 
 \H(a;x)\H(b,c;x) &=& \H(a,b,c;x) + \H(b,a,c;x) + \H(b,c,a;x) \ ; 
\label{eq:halgebraex} \end{eqnarray} 
more complicated cases are immediately established recursively (all the above 
formulae can indeed easily be checked by differentiating, repeatedly when 
needed, with respect to $x$). 
\par 
After this digression, let us come back to Eq.(\ref{eq:thesol}), which gives 
the solution as an indefinite integral. 
The original $n=-2$ integrand consists 
of powers of $x'$, $\ 1/(1-x')$ and $\ 1/(1+x')$, (without any 
loss of generality $(1-x')$ and $\ (1+x')$ can be taken to occur only as 
negative powers), which are present in all the factors (see the first 
of Eq.s(\ref{eq:exp-2-1}) for the definition of $R^{(-2)}(x')$), and of 
the two \hpl's of weight 1, $\H(1;x') = -\ln(1-x')$ and 
$\ \H(-1;x') = \ln(1+x')$, which occur in $\phi_2(x')$, Eq.(\ref{eq:homosol}). 
The terms containing powers of the rational factors different from the 
mere inverse, \ie different from the 
first powers of $1/x'$, $\ 1/(1-x')$ and $\ 1/(1+x')$, can be integrated 
by parts; in so doing they give rise at most to derivatives of the occurring 
\hpl's, \ie again products of the three rational factors 
$x'$, $\ 1/(1-x')$ and $\ 1/(1+x')$, and, 
according to Eq.s(\ref{eq:dxHw}, \ref{eq:defdHw1}), of \hpl's of 
lower weight if the \hpl's in the integrand have weight greater than 1 
(as it is in the general case $n>-2$). The result can be partial 
fractioned, and the integration by parts repeated until all powers of 
the rational fractions different from the mere first power are processed. 
The terms with the mere inverse of the three rational fractions and 
\hpl's, finally, can be integrated at once according to 
Eq.(\ref{eq:intdxHw}), giving rise to \hpl's of higher weight. 
The integration constants are then fixed by 
imposing the $x\to0^+$ behaviour~(\ref{eq:Phiexpld}). 
It is clear that the argument applies to any of the subsequent 
integrations as well, recursively in $n$. \par 
For $n=-2$ the explicit calculation gives $ \Phi^{(-2)}_1 = 0 $, 
$ \Phi^{(-2)}_2 = 0 $ and 
\begin{equation} 
 \Phi^{(-2)}(x) = - \ \frac{2 + x^2}{8} \ . 
\label{eq:Phi-2} 
\end{equation} 
Note that, in this case, all the \hpl's cancel out from the final 
result. 
%%%%%%%%%%%%%%%%%%%%%%%%%%%%%%%%%%%%%%%%%%%%%%%%%%%%%%%%%%%%% 
% which is of course in agreement with the general mass case 
% given in the literature~\cite{sreq}. 
%%%%%%%%%%%%%%%%%%%%%%%%%%%%%%%%%%%%%%%%%%%%%%%%%%%%%%%%%%%%% 
\par 
One can then proceed to $n=-1$; the term $\ln{x}$ appearing in the 
inhomogeneous part of the second of Eq.s(\ref{eq:exp-2-1}) can be written 
as $H(0;x)$, see (\ref{eq:defw1}), the products of the \hpl's already 
present in $R^{(-1)}(x')$ and the \hpl's of the $\phi_i(x')$ can be 
rewritten as combinations of single \hpl's of suitable indices according 
to~\ref{eq:halgebra}, the integration performed as explained above -- 
and so on for the higher values of $n$, up to any required order. 
An explicit calculation gives for the integration constants 
$ \Phi^{(-1)}_1 = -5/64 $, 
$ \Phi^{(0)}_1 = 29/256 $, $ \Phi^{(1)}_1 = - 107/1024 $, 
$ \Phi^{(2)}_1 = 185/4096 $, $ \Phi^{(3)}_1 = - 1285/16384 $, 
$ \Phi^{(4)}_1 = - 18991/65536 $, $ \Phi^{(5)}_1 = 164173/262144 $, 
while, for any $n$, $ \Phi^{(n)}_2 = 0 $. 
The first $\Phi^{(n)}(x) $ are 
\begin{equation} 
 \Phi^{(-1)}(x) = 
       \frac{3}{8}
       + x^2 \left(
            \frac{5}{32}
          - \frac{1}{4} H(0;x)
          \right)  
 \ , 
\label{eq:Phi-1} 
\end{equation} 
\begin{eqnarray} 
 \Phi^{(0)}(x) 
     & = &
          - \frac{3}{8}
          - \frac{1}{8} H(0;x)
        + x^2 \left(
          - \frac{11}{128}
          + \frac{5}{16} H(0;x) 
          - \frac{1}{8} H(0;x) H(0;x) 
                    \right) \nonumber \\ 
     &   &
         - \frac{(1-x^2)^2}{8 x^2} 
         \biggl( 
              H(0;x) H(-1;x) 
            - H(0;x) H(1;x)
            - H(0,-1;x)
            + H(0,1;x) 
         \biggr) \ ,
\label{eq:Phi0} 
\end{eqnarray} 
\begin{eqnarray} 
 \Phi^{(1)}(x)  &=& \frac{(1-x^2)^2}{32 \ x^2} \times \nonumber \\ 
                & & 
                    \left\{ \frac{}{} 
                     \ 5 \left[ \frac{}{}
                                  H(0;x)H(-1;x) 
                                - H(0;x)H(1;x) 
                                - H(0,-1;x) 
                                + H(0,1;x)  
                            \right] \right. \nonumber \\
                & & 
                     + \ 2 \left[ \frac{}{}
                             H(0;x)H(0;x)H(1;x) 
                           - H(0;x)H(0;x)H(-1;x)
                           - H(0;x)H(-1;x)H(-1;x)
                          \right.
                     \nonumber\\ 
                & & \qquad 
                    \left. \frac{}{}
                           + 4 H(0;x)H(-1;x)H(1;x)
                           - H(0;x)H(1;x)H(1;x)
                    \right] \nonumber \\
                & &  
                     + \ 4 \left[ \frac{}{}  
                     H(0,-1;x)H(-1;x) + 2 H(0,-1;x)H(0;x) - 2 H(0,-1;x)H(1;x) 
                           \right] \nonumber\\
                & & 
                     + \ 4 \left[ \frac{}{}  
                     H(0,1;x)H(1;x) - 2 H(0,1;x)H(-1;x) - 2 H(0,1;x)H(0;x) 
                           \right]  \nonumber\\
                & & + \ 8 \left[
                      H(0,-1,1;x) 
                    + H(0,1,-1;x)
                    - \frac{3}{2} H(0,0,-1;x) 
                    + \frac{3}{2} H(0,0,1;x)
                      \right.  \nonumber\\
                & & \left. \qquad
                    - \frac{1}{2} H(0,-1,-1;x)
                    - \frac{1}{2} H(0,1,1;x)
                   \right] \left. \frac{}{} \right\} \nonumber\\ 
                & + &
                     \frac{(1+x)^2}{x}
                       \left[ \frac{}{} 
                       H(0,-1;x) - H(0;x)H(-1;x) 
                       \right] \nonumber\\ 
                & + &
                     \frac{(1-x)^2}{x}
                       \left[ \frac{}{} 
                       H(0,1;x) - H(0;x)H(1;x) 
                       \right] \nonumber\\
                & - & \frac{x^2}{8} 
                     \left[ 
                      \frac{55}{64} + \frac{11}{8} H(0;x) 
                      - \frac{5}{4} H(0;x)H(0;x) 
                      + \frac{1}{3} H(0;x)H(0;x)H(0;x)
                     \right] \nonumber \\ 
                & + &  \frac{15}{64} + \frac{13}{32}H(0;x) 
                      - \frac{1}{16} H(0;x)H(0;x)  \ .
\label{eq:Phi1} 
\end{eqnarray} 
Eq.s(\ref{eq:Phi-2},\ref{eq:Phi-1}) 
are of course in agreement with the general mass case 
(see for instance ~\cite{sreq},\cite{BDU}); 
Eq.(\ref{eq:Phi0}) reproduces Eq.(25) of~\cite{BDU}, while 
Eq.(\ref{eq:Phi1}) is new in the literature. 
As matter of fact we evaluated explicitly, 
by means of an integration routine written in {\tt FORM}~\cite{FORM}, all 
the $\Phi^{(n)}(x)$ up to $n=5$ included, which is found to involve 
\hpl's of weight ${\bf 7}$. The explicit expressions obtained are however 
too long to be listed here. It is found that, in general, 
$\Phi^{(n)}(x)$ involves \hpl's of weight $(n+2)$. 
\par 
The values of the \hpl's of argument equal to 1 are known ~\cite{VermTab}; 
we can then take our explicit analytic results for the $\Phi^{(n)}(x)$ 
up to $n=5$ and evaluate them at $x=1$ by using ~\cite{VermTab}, 
obtaining the analytical on-shell value of the sunrise amplitude 
Eq.~\ref{eq:F0part} in the equal mass case limit up to the 5th 
order in the $(d-4)$ expansion. At $x=1$ the terms of highest weight 
disappear from $\Phi^{(n)}(x)$, due to the general 
structure of the solution, Eq.(\ref{eq:homosol},\ref{eq:thesol}), 
and the coefficient of $(d-4)^n$ involves constants up to weight 
$n+1$ only. The result reads 
\begin{eqnarray}
\Phi(d,x=1) 
    & = & 
       - \frac{3}{8} \frac{1}{(d-4)^2} 
       + \frac{17}{32} \frac{1}{(d-4)} 
       - \frac{59}{128}
       + (d-4) \left(
            \frac{65}{512}
          + \frac{1}{24} \pi^2
          \right) \nonumber \\ 
    &   &
       + (d-4)^2 \left(
            \frac{1117}{2048}
          - \frac{13}{96} \pi^2
          + \frac{1}{8} \pi^2 \ln2
          - \frac{7}{16} \zeta(3) 
          \right) \nonumber \\ 
    &   & 
       + (d-4)^3 \left(
          - \frac{13783}{8192}
          + \frac{115}{384} \pi^2
          - \frac{13}{32} \pi^2 \ln2
          + \frac{91}{64} \zeta(3)
          + \frac{1}{8} \pi^2 \ln^2 2
          - \frac{31}{2880} \pi^4
          + \frac{1}{16} \ln^4 2
          + \frac{3}{2} a_4
          \right) \nonumber \\ 
    &   &
       + (d-4)^4  \left(
            \frac{114181}{32768}
          - \frac{865}{1536} \pi^2   
          + \frac{115}{128} \pi^2 \ln2
          - \frac{805}{256} \zeta(3)
          - \frac{13}{32} \pi^2 \ln^2 2
          + \frac{403}{11520} \pi^4
          - \frac{13}{64} \ln^4 2
          \right. \nonumber \\
    &   & 
        \left. \qquad \qquad \quad 
          - \frac{39}{8} a_4
          - \frac{31}{960} \pi^4 \ln2
          + \frac{1}{8} \pi^2 \ln^3 2
          + \frac{5}{96} \pi^2 \zeta(3)
          + \frac{3}{80} \ln^5 2
          - \frac{9}{2} a_5 
          + \frac{465}{128} \zeta(5)
          \right) \nonumber \\ 
    &   &
         + (d-4)^5 \left(  
          - \frac{820495}{131072}
          + \frac{5971}{6144} \pi^2
          - \frac{865}{512} \pi^2 \ln2
          + \frac{6055}{1024} \zeta(3)
          + \frac{115}{128} \pi^2 \ln^2 2
          - \frac{713}{9216} \pi^4     
          + \frac{115}{256} \ln^4 2
          \right. \nonumber \\
    &   & 
          \qquad \qquad \quad
          + \frac{345}{32} a_4
          - \frac{65}{384} \pi^2 \zeta(3)
          - \frac{13}{32} \pi^2 \ln^3 2  
          + \frac{403}{3840} \pi^4 \ln2
          - \frac{39}{320} \ln^5 2    
          - \frac{6045}{512} \zeta(5)
          + \frac{117}{8} a_5
         \nonumber \\
    &   & 
          \qquad \qquad \quad
          - \frac{9}{160} \pi^4 \ln^2 2
          + \frac{1}{64} \pi^2 \ln^4 2 
          + \frac{79}{34560} \pi^6  
          + \frac{15}{8} \zeta(3) \ln^3 2 
          - \frac{25}{32} \zeta(3) \pi^2 \ln2 
          + \frac{595}{128} \zeta^2(3)
          \nonumber \\ 
    &   & 
          \qquad \qquad \quad
          \left.
          + \frac{45}{4} \zeta(5) \ln2 
          - \frac{45}{4} a_5 \ln2 
          + \frac{21}{320} \ln^6 2   
          - 9 a_6 
          + \frac{45}{4} b_6 \right) 
          \nonumber \\ 
   &   & 
          + {\mathcal O}((d-4)^6) \ , 
\label{eq:Phiat1} 
\end{eqnarray} 
where we have expressed the constant ${\tt s6}$ appearing in~\cite{VermTab} 
as 
\begin{eqnarray}
 {\tt s6} 
    & = &
          - \frac{1}{6} \pi^2 \zeta(3) \ln2 
          - \frac{1}{72} \pi^2 \ln^4 2
          - \frac{1}{720} \pi^4 \ln^2 2
          + \frac{1}{540} \pi^6
          + 2 \zeta(5) \ln 2 
          \nonumber \\
    &   &
          - 2 a_5 \ln2  
          + \frac{1}{3} \zeta(3) \; \ln^3 2 
          + \frac{1}{120} \ln^6 2
          + \frac{5}{4} \zeta^2 (3)
          - 4 a_6
          + 2 b_6
          \ , 
\end{eqnarray}
and the definitions of all the previous constants in terms 
of \hpl's or Nielsen's polylogarithms (NPl) is given in 
Table ~\ref{tab:const}, together with their numerical values. 
\par 
Our result, Eq.(\ref{eq:Phiat1}), agrees with previous results in the 
literature (apart from obvious normalization factors), such as 
\cite{Krakow}, which goes up the $(d-4)^3$ of present work, and 
\cite{Broadhurst} which arrives at $(d-4)^4$. 
\par 
Numerically, Eq.(\ref{eq:Phiat1}) gives 
\begin{eqnarray}
\Phi(d,x=1) 
        & = & 
            - 0.37500000000000000000 \frac{1}{(d-4)^2}
            + 0.53125000000000000000 \frac{1}{(d-4)}
        \nonumber \\ 
        &   & 
            - 0.46093750000000000000 
            + 0.53818664171204166667 (d-4)
        \nonumber \\ 
        &   & 
            - 0.46186261021407291667 (d-4)^2
            + 0.53810855601624843750 (d-4)^3 
        \nonumber \\ 
            &   & 
            - 0.46190033063946289062 (d-4)^4
            + 0.53809670843016515942 (d-4)^5 
        \nonumber \\ 
            &   &
          + {\mathcal O}((d-4)^6) \ , 
\label{eq:Phiat1_num} 
\end{eqnarray}
in perfect agreement with the numerical results of~\cite{numcal}, 
which moreover gives two further terms in the expansion. 
\par 
As a last curiosity, let us observe that for $d\to4$ Eq.(\ref{eq:Phiat1_num})
can be rewritten as 
\begin{eqnarray}
\Phi(d,x=1) 
        & = & \frac{1}{(d-5)} \left(   
            + 0.37500000000000000000 \frac{1}{(d-4)^2}
            - 0.90625000000000000000 \frac{1}{(d-4)} \right. 
         \nonumber \\ 
        &   & 
           \qquad \qquad \  + 0.99218750000000000000 
                            - 0.99912414171204166667 (d-4)
        \nonumber \\ 
        &   & 
           \qquad \qquad \  + 1.0000492519261145833 (d-4)^2
                            - 0.99997116623032135417 (d-4)^3 
        \nonumber \\ 
            &   &
           \qquad \qquad \  + 1.0000088866557113281 (d-4)^4
                            - 0.99999703906962805005 (d-4)^5 
        \nonumber \\ 
           &   &
        \left. \qquad \qquad \frac{}{} + {\mathcal O}((d-4)^6) \right) \ , 
\label{eq:curious} 
\end{eqnarray}
showing that the coefficients of the powers of $(d-4)^n$ approach quickly 
$(-1)^n$ for increasing $n$. 
\vspace*{0.5truecm} 
\begin{table}[!h]
\begin{center}
\begin{tabular}{|c|c|c|l|}
\hline 
 constant & HPL & NPl & numerical value \\ 
\hline 
\hline 
\hline 
 $ \zeta(3) $ & $ H(0,0,1;1) $ & $ S_{2,1}(1) $ 
              & 1.2020569031595942854 \\
 $ a_4 $      & $ H(0,0,0,1;1/2) $ & $ S_{3,1}(1/2) $ 
              & 0.51747906167389938633 \\
 $ \zeta(5) $ & $ H(0,0,0,0,1;1) $ & $ S_{4,1}(1) $
              & 1.0369277551433699263 \\  
 $ a_5 $      & $ H(0,0,0,0,1;1/2) $   & $ S_{4,1}(1/2) $
              & 0.50840057924226870746 \\
 $ a_6 $      & $ H(0,0,0,0,0,1;1/2) $ & $ S_{5,1}(1/2) $
              & 0.50409539780398855069 \\  
 $ b_6 $      & $ H(0,0,0,0,1,1;1/2) $ & $ S_{4,2}(1/2) $
              & 0.0087230030575968884272  \\  
\hline 
\end{tabular}
\end{center}
\caption{Constants up to weight 6 appearing in the calculation}
\label{tab:const}
\end{table}
%%%%%%%%%%%%%%%%%%%%%%%%%%%%%%%%%%%%%%%%%%%%%%%%%%%% 

\section{Conclusions. } 
\label{sec:Concl} 
\setcounter{equation}{0} 
We have shown that the two-loop self-mass sunrise amplitude, in the 
usual $d$-continuous dimensional regularization and 
particular kinematical configuration corresponding to two equal 
masses $M$ and the external invariant equal to the square of the 
third mass $m$, satisfies a second order differential equation 
in the dimensionless ratio $x=m/M$. 
The equation can be expanded in Laurent series at $d=4$, and solved 
in closed analytical form. As a results, 
all the $n$-th coefficient of the Laurent expansion can be 
expressed in terms of harmonic polylogarithms of 
argument $x$ and weight up to $n+2$. As a by product, we give 
the explicit analytic expressions of the value of the amplitude at 
$x=1$, corresponding to the on-mass-shell sunrise amplitude in the 
equal mass case, up to the $(d-4)^5$ term included. 

\vspace*{1cm}\noindent{\Large {\bf Acknowledgments}} \\ 

\noindent 
PM was supported in part by a grant of INFN (bando 7798/99) 
and in part by a grant of the "Marie Curie Training Site" at 
Karlsruhe University.

%%%%%%%%%%%%%%%%%%%%%%%%%%%%%%%%%%%%%%%%%%%%%%%%%%%% 
 

\begin{thebibliography}{99} 
\def\NC{{\sl Nuovo Cim.}} 
\bibitem{sreq} 
M.~Caffo, H.~Czy{\.z}, S.~Laporta and E.~Remiddi , arXiv:hep-th/9805118, 
\NC {\bf A111}, 365 (1998). 
\bibitem{hpl} 
E.\ Remiddi and J.A.M.\ Vermaseren, arXiv:hep-ph/9905237, 
Int.\ J.\ Mod.\ Phys.\ {\bf A15} (2000) 725.
\bibitem{FORM} J.A.M.\ Vermaseren, Symbolic Manipulation with 
{\tt FORM}, Version 2, CAN, Amsterdam, 1991; \\ 
New features of {\tt FORM}, arXiv:math-ph/0010025. 
\bibitem{BDU} F.\ Berends, D.\ Davydychev and N.\ Ussyukina, 
Phys. Lett. B426 (1998) 95. 
\bibitem{VermTab} J.\ Vermaseren, table available from the 
{\tt URL \\ 
http://www.nikhef.nl/$\sim$form/FORMdistribution/packages/harmpol/index.htlm}. 
\bibitem{Krakow} 
S.\ Laporta, E.\ Remiddi, Acta Physica Polonica B, Vol. {\bf B28} 
(1997) 959. 
\bibitem{Broadhurst} D. Broadhurst, private communication. 
% \bibitem{previous}
% A.V.\ Kotikov, arXiv:hep-ph/0102178.
\bibitem{numcal}
S.\ Laporta, Phys. Lett. B523, 95 (2001), arXiv:hep-ph/0111123. 
See in particular Eq.(14) for the quantity $I_{15}$; in order to match the 
normalization our result must be multiplied by $64/(d-2)$ and then 
expanded in $(4-d)/2$. 
\end{thebibliography}
\end{document}